\documentclass[journal]{IEEEtran}
\usepackage{subfig}

\usepackage{etoolbox}
\usepackage{dtsc-creafig}
\usepackage{notation}

\usepackage{epsfig}
\usepackage{color}
\usepackage{amsmath}
\usepackage{amssymb}
\usepackage{mathabx}
\usepackage{enumerate}
\usepackage{multirow}
\usepackage{bbm}
\usepackage{algorithm}
\usepackage{algorithmic}
\usepackage{lipsum,amsmath,multicol}
\usepackage{stfloats}
\usepackage{arydshln} 
\usepackage{amsfonts}
\usepackage{dsfont}
\usepackage{upgreek}
\usepackage{pgfplots} 

\usepackage{booktabs} 

\usepackage{graphicx}

\usepackage{cite}

\usepackage{acronym}
\acrodef{BER}{bit error rate}
\acrodef{LMMSE}{linear minimum-mean squared-error}
\acrodef{RSS}{received signal strength}
\acrodef{OK}{ordinary kriging}
\acrodef{IDW}{inverse distance weighting}
\acrodef{GP}{Gaussian processes}
\acrodef{CRB}{Cram\'er-Rao bound}
\acrodef{RF}{radio frequency}
\acrodef{EIRP}{equivalent isotropically radiated power}
\acrodef{AWGN}{Additive White Gaussian Noise}
\acrodef{BCRB}{Bayesian Cram\'er-Rao bound}
\acrodef{MSE}{mean squared-error}
\acrodef{GPR}{Gaussian process for regression}
\acrodef{HCRB}{hybrid Cram\'er-Rao bound}
\acrodef{TOA}{time of arrival}
\acrodef{BS}{base station}
\acrodef{OKD}{ordinary Kriging with detrending}
\acrodef{IDW}{inverse distance weighting}

\ifCLASSINFOpdf
\else
\fi

\begin{document}
%
\title{{Recursive Estimation of Dynamic RSS Fields Based on Crowdsourcing and Gaussian Processes}}
%
%
%

\author{Irene~Santos,
        Juan~Jos\'e~Murillo-Fuentes, Petar M. Djuri\'c
\thanks{I. Santos and J. J. Murillo-Fuentes are with the Dept. Teor\'ia de la Se\~nal y Comunicaciones, Universidad de Sevilla, Camino de los Descubrimiento s/n, 41092 Sevilla, Spain. E-mail: {\tt \{irenesantos,murillo\}@us.es}}
\thanks{P. M. Djuri\'c is with the Dept. Electrical and Computer Engineering, Stony Brook University, Stony Brook, NY 11794, USA. E-mail: {\tt petar.djuric@stonybrook.edu}.}
\thanks{This manuscript has been submitted to IEEE Transactions on Signal Processing on May 11, 2018; revised on August 9, 2018 and November 15, 2018; accepted on December 10, 2018.  The work of I. Santos and J.J. Murillo-Fuentes was partially funded by Spanish Government (Ministerio de Econom\'ia y Competitividad TEC2016-78434-C3-2-R) and by the European Union (FEDER). P. M. Djuri\'c was supported by the National Science Foundation under Award CNS 1642965.}}

\maketitle


\begin{abstract}
In this paper, we address the estimation of a time-varying spatial field of received signal strength (RSS) by relying on measurements from randomly placed and not very accurate sensors. We employ a radio propagation model where the path loss exponent and the transmitted power are unknown with Gaussian priors whose hyper-parameters are estimated by applying the empirical Bayes method. We consider the locations of the sensors to be imperfectly known, which entails that they represent another source of error in the model. The propagation model includes shadowing which is considered to be a zero-mean Gaussian process where the correlation of attenuation between two spatial points is quantified by an exponential function of the distance between the points. The location of the transmitter is also unknown and is estimated from the data. We propose to estimate time-varying RSS fields by a recursive Bayesian method and crowdsourcing. The method is based on \ac{GP}, and it produces the joint distribution of the spatial field. Further, it summarizes all the acquired information by keeping the size of the needed memory bounded. We also present the \ac{BCRB} of the estimated parameters. Finally, we illustrate the performance of our method with experimental results on synthetic and real data sets.
\end{abstract}


%
\IEEEpeerreviewmaketitle

\section{Introduction}\LABSEC{intro}
%
%
%
%


\IEEEPARstart{S}{pectrum} sensing has gained significant interest for research due to the rapid growth of wireless communication systems. Its main function is to map the distribution of \ac{RF} signals within a specific area, detecting intruders in a particular spectrum band and/or free channels that are not being used by any user \cite{Li12,Portelinha13}. For this purpose, spectrum sensing relies on measurements from sensors. Current techniques for spectrum management require these measurements to be quite accurate, i.e., the techniques need an expensive infrastructure where sensors are sparsely and strategically deployed. These sensors provide precise measurements of received power and their positions are perfectly known. Due to the cost of this setting, approaches based on measurements obtained by expensive equipment do not scale well and cannot be extended to large areas. An alternative and appealing solution is to exploit {\it crowdsourcing}, where many sensors with low-quality acquire much less accurate measurements \cite{Pfammatter15,Molinari15}. For example, one can use not very accurate measurements obtained by a large number of smart phones, and yet can create a more accurate spectrum map than that obtained by a few sparsely located expensive sensors. 
%
%
%
%

Most of the current spectrum monitoring techniques rely on \ac{RSS} measurements. 
Given the \ac{RSS} values at some known locations, the goal of spectrum sensing is to estimate the field of \ac{RSS} at any location within an area of interest. 
Spectrum monitoring is not the only application where \ac{RSS} measurements play a central role. Others include indoor localization \cite{Pak11,Chang17}, tracking \cite{Dashti15}, distance estimation \cite{Mahapatra16} and distributed asynchronous regression \cite{Garrido15}. 

Methods that use \ac{RSS} measurements for making inference are based on radio propagation path loss models. These models depend on different parameters including the locations of the sensors, the path loss exponent, the transmitted power, and shadowing. 
The more these parameters fit the reality, the more accurate the model is. Furthermore, due to the dynamic nature of the signal propagation, the learning of the model parameters that describe the propagation should allow for their adaptation. 

There are two groups of papers where the propagation loss model is used. 
In one group, the authors consider the path loss exponent to be known and static \cite{Zhang17,Li12,Wen16,Vaghefi13b,Vaghefi13}. In the other,  the exponent is unknown, is possibly dynamic,  and is estimated \cite{Mazuelas09, Li06, Liang16}. Often, when indoor environments are studied, different values of the path loss exponent are assumed and estimated, e.g., with nonlinear least-squares techniques. 
The Bayesian methods allow for taking into account the estimation error of the exponent while estimating other more important unknowns or in making decisions.  

Shadowing is another important notion in these studies. It has been commonly modeled by log-normal distributions. Usually, no spatial correlation due to shadowing fading has been assumed and i.i.d. log-normal distributions with the same variance have been used \cite{Zhang17,Mazuelas09,Li06,Liang16,Vaghefi13b}. However, it is well-known that shadowing effects at different locations can change significantly due to varying propagation conditions. For this reason, the authors of \cite{Wen16} address the correlation between \ac{RSS} measurements and propose to estimate \ac{RSS} at specific locations as the value of nearby sensors. In \cite{Portelinha13,Li12,Ferris06,Vaghefi13} and \cite{Romero17}, a full covariance matrix is introduced to model the spatially correlated shadowing. While in  \cite{Vaghefi13} this matrix is known, in \cite{Ferris06,Li12,Portelinha13} and \cite{Romero17} a multivariate zero-mean Gaussian is used to approximate it. In \cite{Ferris06,Li12} and \cite{Portelinha13}, an exponentially distributed coefficient of correlation is used. 

Most of the approaches proposed in the literature require the {transmitted power} to be available at a central unit where all the measurements are processed \cite{Li12,Portelinha13,Romero17,Vaghefi13,Li06,Liang16}. However, the transmitted power may be unavailable and may change with time \cite{Vaghefi13b}. In this situation, one can estimate the transmitted power  \cite{Vaghefi11,Kim07} or eliminate the dependence  of the propagation model on it by computing \ac{RSS} differences between sensors \cite{Vaghefi11,Lee09,Wang09}. 

We point out that the transmitter and sensor locations also affect the radio propagation model. Approaches in the literature assume that these locations are perfectly known or, if estimated locations are available, they are considered as if they are true. To the best of our knowledge, there are no models that introduce errors in  distances between transmitters and receivers. 
Further, one may have an increasing number of measurements at different locations at each time instant, and this can increase considerably the complexity and size of the system \cite{Perez13}. In our paper, we address measurements by sensors whose locations are only approximately known, and we consider a scenario with a large number of measurements that is due to crowdsourcing.

The {variability} in the \ac{RSS} and/or the parameters that affect these measurements are rarely discussed. In \cite{Romero17}, an adaptive kernel-based approach is proposed, but there is no discussion on the variations of RSS in practice. In \cite{Chapre13}, some results on \ac{RSS} measurements are reported for a WLAN with  time, while in \cite{Chang17} some variations over a few seconds and days are reported to emphasize the need for updating the fingerprint for localization purposes. In both cases, indoor WiFi signals are considered. 

In this paper, we propose a \ac{GP}-based approach, where the uncertainties in the positions of the sensors {and the transmitter} as well as the correlation due to shadowing effects are included in the model. We also deal with an unknown path loss exponent and transmitted power by modeling them as Gaussian random variables whose hyper-parameters are learned according to the empirical Bayesian approach \cite{Carlin00}. Preliminary results of this approach for static fields can be found in  \cite{Santos17b}. 
We cope with the temporal changes of the \ac{RSS} measurements from some sensors by using a recursive technique where the solution is updated with the arrival of new data. This solution also allows for adaptation to accommodate changes in both propagation and movement of sensors, as showed in \cite{Santos17c} with a synthetic dataset. 

As in previous approaches \cite{Santos17b,Santos17c}, we fix the locations where estimates of RSS are needed. These locations are represented as {\it nodes} of a predefined grid.
The computational complexity is fixed and determined by the number of nodes. This differs from other approaches, e.g., in  \cite{Perez13}, where new available measurements are included sequentially in the system, thereby increasing its size and complexity.
Furthermore, in our work the estimates of the field at the nodes serve as priors for the field estimates at the next time instant when new measurements are acquired. In this way, we make the approach scalable. To account for the time-variations of the field, we include a forgetting factor in the method which determines the relevance of the previous and current data. We also find the \ac{BCRB} of our estimates, where we rely on some recent results on bounds for GPs \cite{Waagberg17}.  Finally, we demonstrate the performance of the method on synthetic  datasets. We compare the performance with previous static approaches based on GPs \cite{Santos17b} and interpolation techniques \cite{Molinari15}.

The paper is organized as follows. We introduce the notation and the model in \SEC{mod}. In \SEC{alphaP}, we explain how to estimate the location of the transmitter and how to find the Gaussian distribution of the path loss exponent and transmitted power. We develop the GP implementation for RSS estimation at static fields in \SEC{GP}. The \ac{BCRB} is presented in \SEC{BCRB}, and the method for estimating time-varying fields in \SEC{recursiveGP}. We provide simulation results in \SEC{sim} \newt{and results from real data in \SEC{New}}. Our concluding remarks are given in \SEC{conclusion}.

\section{System Model}\LABSEC{mod}

In this section we propose a model for the measurements of the sensors. This model will be used to generate synthetic data and to propose an estimation algorithm. We focus on the estimation of \ac{RSS} at $M$ fixed grid nodes with perfect known locations given by $\matr{X}_g=[\vect{x}_{g_1}\cdots \vect{x}_{g_M}]\trs\in \mathbb{R}^{M\times2}$. We denote this estimated field at instant $t$ as ${\bf f}_g^{[t]}\in{\mathbb R}^M$. We will use the superscript $^{[t]}$ to indicate that the variable is a function of time.   
At time $t$, we consider that $N$ random low-cost sensors\footnote{Note that we have removed the superscript $t$ from $N$ for easier reading. } 
appear within the area at locations $\matr{X}^{[t]}=[\vect{x}_1^{[t]} \cdots \vect{x}_N^{[t]}]\trs\in \mathbb{R}^{N\times2}$. We denote the distance between sensors $i$ and $j$ at time $t$ by $d_{ij}^{[t]}$. 
The distance between them is computed from their locations $\vect{x}_i$ and $\vect{x}_j$ by 
\begin{align}
\label{eq:distance}
d_{ij}^{[t]}=\sqrt{(\vect{x}_i^{[t]}-\vect{x}_j^{[t]})\trs(\vect{x}_i^{[t]}-\vect{x}_j^{[t]})}. 
\end{align}

\newt{Given a single transmitter with \ac{EIRP} $P^{[t]}$ (in dBm) at location $\vect{x}_0\in\mathbb{R}^{2\times1}$, the distances between the sensors and the transmitter and the grid nodes and the transmitter are denoted by $\vect{d}^{[t]}=[d_1^{[t]}\cdots d_N^{[t]}]\trs$ and  $\vect{d}_g=[d_{g_1}\cdots d_{g_M}]\trs$, respectively,} and they are defined analogously to \eqref{eq:distance}. All the distances are measured in meters. The sensors also report their measurements of \ac{RSS} (in dBm), and they are denoted by  $\vect{z}^{[t]}=[z_1^{[t]} \cdots z_N^{[t]}]^\top$. Assuming a log-normal path loss model, we express the \ac{RSS} as
\begin{align} \LABEQ{measurementsNoNoise}
\vect{z}^{[t]}&= {\sf \vect{1}} P^{[t]} - 10\alpha^{[t]} \log_{10}(\vect{d}^{[t]})  + \vect{v}^{[t]} + \vect{w}^{[t]},
\end{align}
where ${\sf \vect{1}}$ is an $N\times 1 $ vector of 1's, $\alpha^{[t]}$ is the path loss exponent at time $t$, $\vect{v}^{[t]}$ is an attenuation due to shadowing effects and modeled according to 
\begin{align}
\vect{v}^{[t]}&\sim  \gauss{\vect{v}^{[t]}}{\vect{0}}{\matr{\Sigma}_v^{[t]}},
\label{eq:v}
\end{align}
where the notation $\gauss{\vect{v}^{[t]}}{\vect{0}}{\matr{\Sigma}_v^{[t]}}$ signifies that $\vect{v}^{[t]}$ has a Gaussian distribution with a mean vector $\vect{0}$ and a covariance matrix $\matr{\Sigma}_v^{[t]}$.  
The covariance matrix $\matr{\Sigma}_v^{[t]}$ is comprised of elements given by 
\begin{align}\LABEQ{correlation}
{\rm Cov}\left(v_i^{[t]} v_j^{[t]}\right) & =  {\sigma_v^2} {\rm exp}\left(-{d}_{ij}^{[t]}/D_{\rm corr}\right),
\end{align}
where $D_{\rm corr}$ is a parameter that models the correlation in the measurements, 
and $\vect{w}^{[t]}$ is some unrelated additive noise, 
\begin{align}
\vect{w}^{[t]}&\sim  \gauss{\vect{w}^{[t]}}{\vect{0}}{\sigma_w^2 \matr{I}_N}.
\label{eq:w}
\end{align}

We reiterate that the sensors do not perfectly know their locations (or distances) and instead they only have the estimates of the distances. More specifically, based on the estimates of their positions, $\widehat{\matr{X}}^{[t]}=[\widehat{\vect{x}}_1^{[t]}\cdots \widehat{\vect{x}}_N^{[t]}]$, one can obtain the estimates of their distances from the transmitter, $\widehat{\vect{d}}^{[t]}=[\widehat{d}_1^{[t]} \cdots \widehat{d}_N^{[t]}]$ (see below). We model the estimated distance between the $i$th sensor and the transmitter according to\newt{\footnote{{This model holds if the transmitter position is known. If the position is estimated from the samples, the model holds asymptotically as $N$ grows.}}}  
\begin{align}
\label{location}
\widehat{{d}}_i^{[t]} = d_i^{[t]}+\epsilon_d &\sim  \gauss{\widehat{{d}}_i^{[t]}}{{d}_i^{[t]}}{\sigma_d^2},
\end{align} 
where $\sigma_d^2$ is known. To reflect the uncertainty in $d_i^{[t]}$, we modify \EQ{measurementsNoNoise} to 
\begin{align}\label{measurements}
\vect{z}^{[t]}&= {\sf \vect{1}} P^{[t]} - \widehat{\vect{q}}^{[t]} \alpha^{[t]} + \vect{u}^{[t]} + \vect{v}^{[t]} + \vect{w}^{[t]}, 
\end{align}
where
\begin{align}
\widehat{\vect{q}}^{[t]}&= \left[10\log_{10}(\widehat{d}_1^{\;[t]}) \cdots 10\log_{10}(\widehat{d}_N^{\;[t]})\right]^\top,
\end{align}
and $\vect{u}^{[t]}$ is the error that reflects the imprecisely known location of the sensors, 
\begin{align}\LABEQ{erroru}
\vect{u}^{[t]}&\sim  \gauss{\vect{u}^{[t]}}{\vect{0}}{(\rho_u^{[t]})^2 \widehat{\matr{D}}^{[t]}},
\end{align}
with $\rho_u^{[t]}=10\alpha^{[t]}\sigma_d\log_{10}(e)$ (see the Appendix; $\rho_u^{[t]}$ is expressed in mdB) 
and $\widehat{\matr{D}}^{[t]}={\rm diag}\left\{1/\widehat{d}_1^{2[t]}\cdots 1/\widehat{d}_N^{2[t]}\right\}$. 

In the rest of the paper, we assume that both $P^{[t]}$ and $\alpha^{[t]}$ are unknown, where the EIRP can change with time and the path loss exponent is constant. We consider that both variables are Gaussian distributed as explained in \SEC{alphaP}. 
{We also assume that the location of the transmitter, $\vect{x}_0$, is unknown and static}.  
We let the remaining parameters to be known with values \newt{$\rho_u=200$ mdB ($\sigma_d=13.16$ m)}
and $\sigma_w=\sqrt{7}$ dB. \newt{These values can be previously estimated for the scenario at hand, see, e.g., \cite{Chakraborty2018,Chang17, Chapre13,Talvitie13,Wen16,Zhang17}, or one can extend the proposed empirical Bayes method to estimate them}. Given $\vect{z}^{[t]}$, $\widehat{\matr{X}}^{[t]}$, $\widehat{\vect{d}}^{[t]}$, 
the model in \eqref{measurements}, and all the assumptions, we need to estimate the RSS, $\vect{f}_g^{[t]}$, at the $M$ grid locations, $\matr{X}_g$. \newt{ We point out that $\sigma_d$ is assumed known so that we can generate samples in our simulations, and that it is not needed in the inference stage.}


Note that the model in \eqref{measurements} can be simplified by ignoring the estimation errors of the distances. In that case, we remove   
$\vect{u}^{[t]}$ from the equation and work with 
\begin{align}
\label{mod3}
\hspace{-0.7cm}\vect{z}^{[t]}&= {\sf \vect{1}} P^{[t]} - 10\alpha^{[t]} \log_{10}(\widehat{\vect{d}}^{[t]})  + \vect{v}^{[t]} + \vect{w}^{[t]}. 
\end{align}
In \SSEC{simLoc}, we will compare the performances of the models based on 
\eqref{measurements} and \eqref{mod3}, respectively.

\subsection{Errors of sensor locations}\LABSSEC{errorloc}

In \FIG{errord}, we plotted the measured power at distance $d_i$ from the transmitter as described by \newt{the mean of \eqref{measurements}, i.e., $E(z_i)=P -10\alpha \log_{10} d_i$ for $P = 0$ dBm,}
%
$\alpha = 3.5$ and distances from $0$ to $250$ m when the distance is known correctly (red solid line). The dotted blue curve shows the measurements according to the model when the incorrect distance has an error of $+10$ m, i.e., $\widehat{d}_i=d_i+10$ m. 
We also included two more curves, $-10\alpha \log_{10} \widehat{d}_i\pm\rho_u/\widehat{d}_i$, where $\rho_u=200$. It can be observed that the error of the model at small distances is quite high. The curves suggest how the incorrect estimate of the sensor's distance to the transmitter affects the estimates of the remaining unknowns. 
\newt{As the sensor approaches the transmitter, the detriment of this error  to the estimation of the unknowns increases (the error is kept fixed in the figure).} 



\begin{figure}[htb]
\begin{center}
\includegraphics[width=3.5in]{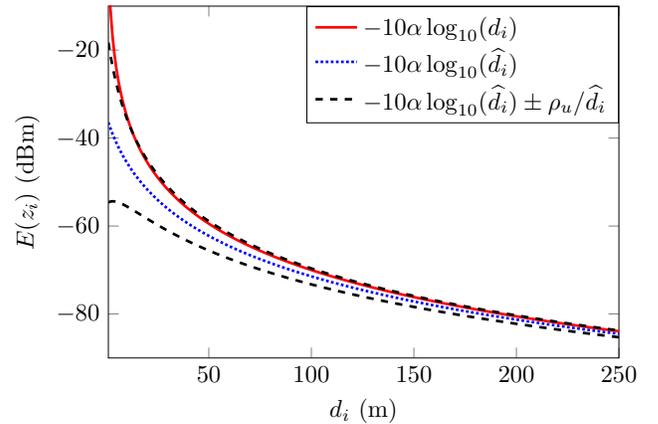}
\caption{The measured powers according to the model in \eqref{measurements} when the distance is known correctly (solid red) or with errors of $+10$ m (dotted blue) and with $\pm$ one standard deviation of the error (dashed black) when $\rho_u=200$ mdB ($\sigma_d=13.16$ m) and $\alpha=3.5$. }
\LABFIG{errord}
\end{center}
\end{figure}

\section{{Estimation of the unknown parameters}}\LABSEC{alphaP}
{In the model given by \eqref{measurements}, we consider that the location of the transmitter, $\vect{x}_0^{\newt{[t]}}$, the path loss exponent, $\alpha^{[t]}$, and the \ac{EIRP}, $P^{[t]}$, are unknown. In the following, we explain how we find the estimates of these parameters. }

\subsection{Estimation of the transmitter location}

We estimate the location of the transmitter from the measurements of RSS at the reporting sensors by the weighted centroid approach \cite{Nurminen17}. \newt{Rather than using just the current RSS measurements, $\vect{z}^{[t]}$, we propose to include all the past information about the transmitter location  as}
\newt{\begin{align}\LABEQ{loc_est}
\widehat{\vect{x}}_0^{[t]}&=\frac{1 }{\hat{w}^{[t]}} \left(\sum_{i=1}^{N} w_{i}^{[t]} \widehat{\vect{x}}_i^{[t]}+{\hat{w}^{[t-1]}}\widehat{\vect{x}}_0^{[t-1]}\right),
\end{align}}

where 
\begin{align}\LABEQ{weights}
w_{i}^{[t]}&={10^{z_{i}^{[t]}/10}}, \\
\hat{w}^{[t]}&=\hat{w}^{[t-1]}+\sum_{i=1}^{N} w_{i}^{[t]}.
\end{align}

Thus,
\begin{align}\label{dist-1-est}
\widehat{d}_{i}^{\newt{[t]}}&=|\widehat{\vect{x}}_i^{\newt{[t]}}-\widehat{\vect{x}}_0^{\newt{[t]}}|,\;\;\;\; i=1, 2, \ldots, N,\\
\label{dist-2-est}
\widehat{d}_{g_i}^{\newt{[t]}}&=|\vect{x}_{g_i}-\widehat{\vect{x}}_0^{\newt{[t]}}|,\;\;\; i=1, 2, \ldots, M. 
\end{align}

We note that the error in estimating $\widehat{\bf x}_0^{\newt{[t]}}$ propagates in the estimates of the distances defined in \eqref{dist-1-est} and \eqref{dist-2-est}.

\subsection{Bayesian estimation of $\alpha^{[\MakeLowercase{t}]}$ and $P^{[\MakeLowercase{t}]}$}

We assume that both variables, $\alpha^{[\MakeLowercase{t}]}$ and $P^{[\MakeLowercase{t}]}$, are independent and Gaussian distributed, i.e.,  
\begin{align}
\label{eq:P}
p(P^{[t]}|\boldsymbol{\theta}_P)&=\gauss{P^{[t]}}{\mu_P^{[t]}}{\sigma_P^{2\,[t]}}, \\
\label{eq:al}
p(\alpha^{[t]}|\boldsymbol{\theta}_\alpha)&=\gauss{\alpha^{[t]}}{\mu_\alpha^{[t]}}{\sigma_\alpha^{2\,[t]}}, 
 \end{align}
 where the hyper-parameters $\boldsymbol{\theta}^{[t]}=[\mu_\alpha^{[t]},\sigma^{[t]}_\alpha,\mu_P^{[t]},\sigma^{[t]}_P]$ have to be estimated from the current set of measurements, $\vect{z}^{[t]}$. To ease the reading, we remove the superscript ${[t]}$ in the remaining of this section. 
 
The joint posterior distribution of $\alpha$ and $P$ given a set of measurements, $\vect{z}$, from sensors with given positions, $\widehat{\matr{X}}$, can be computed by using the Bayes' rule, or
\begin{align}
p(\alpha,P|\vect{z},\widehat{\matr{X}},\boldsymbol{\theta}) &= \frac{p(\vect{z}|\alpha,P,\widehat{\matr{X}}) p(\alpha|\boldsymbol{\theta})p(P|\boldsymbol{\theta})}{p(\vect{z}|\widehat{\matr{X}},\boldsymbol{\theta})},
\end{align}
where  $p(\vect{z}|{\alpha,P},\widehat{\matr{X}})$ is a Gaussian distribution of $\vect{z}$, $p(\vect{z}|{\alpha,P},\widehat{\matr{X}})=\gauss{\vect{z}}{\boldsymbol{\mu}_{z|\alpha,P}}{\matr{\Sigma}_{z|\alpha,P}}$, with
\begin{align}
\boldsymbol{\mu}_{z|\alpha,P}&=\vect{1}P-\widehat{\vect{q}}{\alpha}, \\
\matr{\Sigma}_{z|\alpha,P}&=\newt{\rho_u^2}\widehat{\matr{D}}+\matr{\Sigma}_v+\sigma_w^2 \matr{I}_N. 
\end{align}
%
The marginalized distribution of the measurements, $p(\vect{z}|\widehat{\matr{X}},\boldsymbol{\theta})$, can be computed by
\begin{align}\LABEQ{marginalaP}
p(\vect{z}|\widehat{\matr{X}},\boldsymbol{\theta})&=\int p(\vect{z}|\alpha,P,\widehat{\matr{X}}) p(\alpha|\boldsymbol{\theta})p(P|\boldsymbol{\theta}) \df{\alpha}\df{P},
\end{align}
which is also a Gaussian distribution \cite{Bishop06}, $p(\vect{z}|\widehat{\matr{X}},\boldsymbol{\theta})\sim\gauss{\vect{z}}{\boldsymbol{\mu}_z}{\matr{\Sigma}_z}$, with
\begin{align}
\boldsymbol{\mu}_z&=\vect{\rm \vect{1}}\mu_P -  \widehat{\vect{q}}\mu_\alpha, \LABEQ{meanz} \\
\matr{\Sigma}_z&=\matr{\Sigma}_{z|\alpha,P}+  \sigma^2_\alpha \widehat{\matr{Q}} + \sigma_P^2 \matr{J}_N, \LABEQ{varz} 
\end{align}
where $\widehat{\matr{Q}}=\widehat{\vect{q}}  \widehat{\vect{q}}\trs$ and $\matr{J}_N$ is an $N\times N$ matrix with all elements equal to one.

We can obtain estimates of the hyper-parameters in \eqref{eq:P} and \eqref{eq:al} from the data by applying the empirical Bayes method \cite{Carlin00,Santos17b}. 
More specifically, we approximate $\boldsymbol{\mu}_z$ with $\vect{z}$ and solve the system of equations in \EQ{meanz} to obtain the estimated hyper-parameters of the mean as the following optimization problem:
\begin{align}\LABEQ{meanPalpha}
\begin{bmatrix}
\widehat{\mu}_P\\
\widehat{\mu}_\alpha
\end{bmatrix}
&=
\min_{\mu_P,\mu_\alpha} 
\norm{
\begin{bmatrix}
{\sf \vect{1}} & -\widehat{\vect{q}}
\end{bmatrix}
\begin{bmatrix}
{\mu}_P\\
{\mu}_\alpha
\end{bmatrix}
- \vect{z}
}^2
\; \;\; \mbox{subject to } \mu_\alpha\geq2.
\end{align}
{We recall that the sensors close to the transmitter introduce higher errors in the estimation of $\alpha$ and $P$ than the sensors that are further away, as already explained in \SSEC{errorloc}. For this reason, we used a weighting scheme to account for this in the sum of the squares in \EQ{meanPalpha} as follows:
\begin{align}\LABEQ{meanPalphaChi}
\begin{bmatrix}
\widehat{\mu}_P\\
\widehat{\mu}_\alpha
\end{bmatrix}
&=
\min_{\mu_P,\mu_\alpha} \sum_{i=1}^N \left(\frac{\mu_P -q_i\mu_\alpha -z_i}{\newt{1/\widehat{d}_i}}\right)^2
\; \;\; \mbox{subject to } \mu_\alpha\geq2 \nonumber \\
&=
\min_{\mu_P,\mu_\alpha}
\norm{
\begin{bmatrix}
{\sqrt{\widehat{\matr{D}}}\inv\sf \vect{1}} & -\sqrt{\widehat{\matr{D}}}\inv\widehat{\vect{q}}
\end{bmatrix}
\begin{bmatrix}
{\mu}_P\\
{\mu}_\alpha
\end{bmatrix}
- \sqrt{\widehat{\matr{D}}}\inv\vect{z}
}^2.
\end{align}
}

Next, we approximate $\matr{\Sigma}_z$ with $(\vect{z}-\widehat{\boldsymbol{\mu}}_z)(\vect{z}-\widehat{\boldsymbol{\mu}}_z)\trs$, where $\widehat{\boldsymbol{\mu}}_z=\vect{\rm \vect{1}}\widehat{\mu}_P -  \widehat{\vect{q}}\widehat{\mu}_\alpha$, and obtain the estimates of the hyper-parameters of the variance from \EQ{varz} by solving
\begin{align}
\begin{bmatrix}
\widehat{\sigma}^{2}_P\\
\widehat{\sigma}^2_\alpha
\end{bmatrix}
&=
\min_{\sigma_P^2,\sigma_\alpha^2}
\norm{
\begin{bmatrix}
{\sf \vect{1}} & \vect{b}
\end{bmatrix}
\begin{bmatrix}
{\sigma}^2_P\\
{\sigma}^2_\alpha
\end{bmatrix}
- \vect{a}
}^2
\; \;\; \mbox{subject to } \sigma_\alpha^2,\sigma_P^2\geq0,
\LABEQ{varPalpha}
\end{align}
where $\vect{a}$ and $\vect{b}$ are vectors formed by the diagonal elements of $\matr{A}=(\vect{z}-\widehat{\boldsymbol{\mu}}_\vect{z}) (\vect{z}- \widehat{\boldsymbol{\mu}}_\vect{z})\trs- \boldsymbol{\Sigma}_{\vect{z}|\alpha,P}$ 
and $\widehat{\matr{Q}}$, respectively. 
{Note that the estimation of the variances in \EQ{varPalpha} requires knowledge of $\matr{\Sigma}_v$, i.e., the values of the parameters $D_{corr}$ and $\sigma_v^2$. When these parameters are not available, the previous variances can be learned from the GP, as proposed in the next section. }

\subsection{\newt{Refinement of $\vect{x}_0^{[t]}$, $\alpha^{[t]}$ and $P^{[t]}$}}\LABSSEC{refine}

\newt{The previous estimation of parameters $\vect{x}_0^{[t]}$, $\alpha^{[t]}$ and $P^{[t]}$ can be further refined by using some of the ideas proposed in \cite{Suwansantisuk15}. First, we initialize the location of the transmitter as in \EQ{loc_est}. Then, we estimate the path loss exponent and the transmitted power as in \EQ{meanPalphaChi}. Next, we refine the position of the transmitter as \cite{Suwansantisuk15}, i.e., 
\begin{align}\LABEQ{reest_xo}
\widehat{\vect{x}}_0^{[t]}  &= \min_{\vect{x}_0} \sum_{i=1}^N \left( z_i^{[t]}-\widehat{\mu}_P^{[t]} + \right.\nonumber
\\ 
&\left.+10\widehat{\mu}_\alpha^{[t]} \log_{10}\sqrt{(\vect{x}_i^{[t]}-\vect{x}_0)\trs(\vect{x}_i^{[t]}-\vect{x}_0)} \right)^2
\end{align}
Finally, this refined estimated position of the transmitter is used to obtain new estimates of $\widehat{\mu}_\alpha^{[t]}$ and $\widehat{\mu}_P^{[t]}$ by means of \EQ{meanPalphaChi}. }

\section{GP for regression in static fields}\LABSEC{GP}

In this section, we present an approach to estimate the spatial distribution of \ac{RSS} 
based on a \ac{GPR}. 
We have a set of RSS measurements  $\vect{z}^{[t]}$, obtained at locations $\widehat{\matr{X}}^{[t]}$, and we want to estimate the \ac{RSS} levels $\vect{f}_g^{[t]}$ at a set of given nodes. Note that for now we do not use any information from previous time instants. 
An equivalent formulation of 
\eqref{measurements} is given by
\begin{eqnarray}\LABEQ{sysmodGP}
\vect{z}^{[t]}&=&{f}(\widehat{\matr{X}}^{[t]}) + \vect{\noise}^{[t]},
\end{eqnarray}
where 
$\vect{\noise}^{[t]}$ is an additive Gaussian noise vector defined by 
\begin{eqnarray}
\vect{\noise}^{[t]} &=&\vect{u}^{[t]}+\vect{w}^{[t]} \nonumber \\
&\sim& \gauss{\vect{\noise}^{[t]}}{\vect{0}}{\matr{\Sigma}_\noise^{[t]}=\sigma_w^2\matr{I}_N+\rho_u^2\widehat{\matr{D}}^{[t]}},
\end{eqnarray}
and the function $f(\widehat{\matr{X}}^{[t]})={\rm \vect{1}}P^{[t]}-\widehat{\vect{q}}^{[t]}\alpha^{[t]}+\vect{v}^{[t]}$ is modeled as a GP, i.e., 
\begin{equation}
f(\widehat{\matr{X}}^{[t]})\sim\GP{f(\widehat{\matr{X}}^{[t]})}{\vect{m}_{\widehat{\matr{X}}}^{[t]}}{\matr{K}_{\widehat{\matr{X}}}^{[t]}},
\end{equation}
whose 
 mean  and covariance functions are given by
\begin{eqnarray}
\vect{m}_{\widehat{\matr{X}}}^{[t]}&=&\vect{\rm \vect{1}}\widehat{\mu}_P^{[t]} -  \widehat{\vect{q}}^{[t]}\widehat{\mu}_\alpha^{[t]}, \LABEQ{meanfx}\\
\matr{K}_{\widehat{\matr{X}}}^{[t]}&= &\matr{K}(\widehat{\matr{X}}^{[t]},\widehat{\matr{X}}^{[t]}), 
\end{eqnarray}
where $\matr{K}(\widehat{\matr{X}}^{[t]},\widehat{\matr{X}}^{[t]})$ is an $N\times N$ matrix with elements $k(\widehat{\vect{x}}_i^{[t]},\widehat{\vect{x}}_j^{[t]})$ determined by a specific kernel, defined in \SSEC{kernel}, and 
$\widehat{\boldsymbol{\theta}}^{[t]}=[\widehat{\mu}_\alpha^{[t]},\widehat{\sigma}^{[t]}_\alpha,\widehat{\mu}_P^{[t]},\widehat{\sigma}^{[t]}_P,{\widehat{\vect{x}}_0^{[t]}}]$ are the estimated hyper-parameters of the GPR computed as described in \SEC{alphaP}. 
Note that we have not adopted zero-mean functions, as commonly done \cite{Rasmussen06},  because this assumption would violate the model from \eqref{measurements}. {As already discussed in \SEC{alphaP}, the hyper-parameters $\widehat{\sigma}^{[t]}_P$ and $\widehat{\sigma}^{[t]}_\alpha$ can be computed if the values of the parameters $\sigma_v^2$ and $D_{corr}$ are known. If $\sigma_v^2$ and $D_{corr}$ are not known, we propose to estimate them as parameters of the kernel which will be learned from the GP (see \SSEC{kernel}). }





Given a set of training points, ($\vect{z}^{[t]}, \widehat{\matr{X}}^{[t]}$), we want to estimate the spatial field, $\vect{f}_g^{[t]}$, at a set of test points, $\matr{X}_g$, whose prior is distributed according to
\begin{align}
p(\vect{f}_g^{[t]}|\boldsymbol{\theta}^{[t]})&= \gauss{\vect{f}_g^{[t]}}{\vect{m}_{{\matr{X}_g}}^{[t]}}{\matr{K}_{{\matr{X}}_g}^{[t]}},
\end{align}
where
\begin{align}
\vect{m}_{{\matr{X}_g}}^{[t]}&=\vect{\rm \vect{1}}\widehat{\mu}_P^{[t]} -  \vect{q}_g\widehat{\mu}_\alpha^{[t]},\LABEQ{muXg} \\
\matr{K}_{{\matr{X}}_g}^{[t]}&=\matr{K}({\matr{X}_g},{\matr{X}_g}), \\
\vect{q}_g& = [10\log_{10}(\widehat{d}_{g_1})\cdots 10\log_{10}(\widehat{d}_{g_M})]\trs.
\end{align}
We note that in the last equation we have a hat symbol above ${d}_{g_i}$ because even though we know the exact locations of the nodes where we estimate the RSS, we do not know the exact location of the transmitter. 

The joint distribution of the training outputs, $\vect{z}^{[t]}$, and the test outputs, $\vect{f}_g^{[t]}$, that fits the model and priors above is 
\begin{align}\LABEQ{jointGP}
\begin{bmatrix}
\vect{z}^{[t]} \\
\vect{f}_g^{[t]}
\end{bmatrix}
\sim
\gauss{\begin{bmatrix}
\vect{z}^{[t]} \\
\vect{f}_g^{[t]}
\end{bmatrix}}{\begin{bmatrix}
\vect{m}_{\widehat{\matr{X}}}^{[t]} \\
\vect{m}_{{\matr{X}_g}}^{[t]}
\end{bmatrix}}{\begin{bmatrix}
\matr{K}_{\widehat{\matr{X}}}^{[t]}+\matr{\Sigma}_\noise^{[t]} & \matr{K}_{{\matr{X}}_g,{\widehat{\matr{X}}}}^{[t]\,\top} \\
\matr{K}_{{\matr{X}}_g,{\widehat{\matr{X}}}}^{[t]} & \matr{K}_{{\matr{X}}_g}^{[t]}
\end{bmatrix}}.
\end{align}
The prediction of the RSS at the desired nodes is presented by their posterior distribution, $\vect{f}_g^{[t]}$. This distribution is obtained by conditioning on the observations, $\vect{z}^{[t]}$, estimated locations of the sensors, $\widehat{\matr{X}}^{[t]}$, and the estimated hyper-parameters, $\widehat{\boldsymbol{\theta}}^{[t]}$, i.e., 
\begin{align}\LABEQ{condunav2}
p\left(\vect{f}_{g}^{[t]}|\matr{X}_{g},{\widehat{\matr{X}}}^{[t]},\vect{z}^{[t]},\widehat{\boldsymbol{\theta}}^{[t]}\right)&= \gauss{\vect{f}_{g}^{[t]}}{\boldsymbol{\mu}_g^{[t]}}{\boldsymbol{\Sigma}_g^{[t]}},
\end{align}
where
\begin{align}\LABEQ{mGP}
\boldsymbol{\mu}_g^{[t]}&=\vect{m}_{\matr{X}_g}^{[t]}\!+\! \matr{K}_{\matr{X}_g,{\widehat{\matr{X}}}}^{[t]}{\left(\matr{C}_{\widehat{\matr{X}}}^{[t]}\right)}\inv \;\!\!\!\!\left(\vect{z}^{[t]}-\vect{m}_{\widehat{\matr{X}}}^{[t]}\right),\\
\boldsymbol{\Sigma}_g^{[t]}&=\matr{K}_{{\matr{X}}_g}^{[t]}\!-\!\matr{K}_{{\matr{X}}_g,{\widehat{\matr{X}}}}^{[t]}{\left(\matr{C}_{\widehat{\matr{X}}}^{[t]}\right)}\inv \matr{K}_{{\matr{X}}_g,{\widehat{\matr{X}}}}^{[t]\,\top},\LABEQ{vGP} \\
\matr{C}_{\widehat{\matr{X}}}^{[t]}&=\matr{K}_{\widehat{\matr{X}}}^{[t]}+\boldsymbol{\Sigma}_\noise^{[t]}. 
\end{align}

We refer to this solution as static GP-based approach (sGP), and it is summarized in \ALG{staticGP}.

\begin{algorithm}[htb]
\begin{algorithmic}
\STATE 
Given a specific time instant $t$ and the RSS measured by sensors ($\vect{z}^{[t]}$) at positions $\widehat{\matr{X}}^{[t]}$: 
\STATE
{\bf1)} Obtain the estimates of the hyper-parameters, 
\begin{eqnarray}
\widehat{\boldsymbol{\theta}}^{[t]}&=&\left[\widehat{\mu}_\alpha^{[t]},\widehat{\sigma}^{[t]}_\alpha,\widehat{\mu}_P^{[t]},\widehat{\sigma}^{[t]}_P,{\widehat{\vect{x}}_0^{[t]}}\right], \nonumber
\end{eqnarray}
by solving \EQ{meanPalpha} and \EQ{varPalpha}. 
\STATE
{\bf2)} Compute the posterior distribution at the grid nodes $p\left(\vect{f}_{d}^{[t]}|\matr{X}_{d},{\widehat{\matr{X}}}^{[t]},\vect{z}^{[t]},\widehat{\boldsymbol{\theta}}^{[t]}\right)$ in \EQ{condunav2}. 
\end{algorithmic}
\caption{Static GP-based approach (sGP) formulation}\LABALG{staticGP}
\end{algorithm}

\subsection{The kernel and its hyper-parameters}\LABSSEC{kernel}

{In implementing the GPR, we need to select the kernel of the GPR which measures the similarity between points and is used to predict values of RSS at the nodes of interest from the measured RSS. We chose to work with a combination of different kernels given by
%
\begin{align}
k(\vect{x}_i,\vect{x}_j) &= {\sigma_k^2} {\rm exp}\left(-\frac{\sqrt{(\vect{x}_i-\vect{x}_j)(\vect{x}_i-\vect{x}_j)\trs}}{2l^2}\right) +  \nonumber \\
&+{\sigma}_\alpha^2\widehat{q}(\vect{x}_i)\widehat{q}(\vect{x}_j) + {\sigma}_P^2,
\end{align}
where $\widehat{q}(\vect{x}_i)=10\log_{10}\sqrt{(\vect{x}_i-\widehat{\vect{x}}_0)(\vect{x}_i-\widehat{\vect{x}}_0)\trs}$. 
The parameters $\sigma_k$, $l$, ${\sigma}_\alpha$ and ${\sigma}_P$ are obtained by the GP by minimizing the log marginal likelihood. }

\newt{The covariance matrix above has three terms. The first term is an exponential kernel, which allows for learning of the field from training data. The second and third terms explain the final uncertainty due to estimation of the path loss exponent and the transmitted power, respectively.}

{Here, we make an important point about the proposed approach. It is based on a detailed description of the system by a mathematical model. The objective of the model is to explain as much of the observed data as possible. This also allows the GP to learn more quickly and to correct for the modeling errors. In other words, GPs, being quite robust, will  compensate for errors in the estimation of the mean term, i.e., in the exponent loss and transmitted power in \EQ{meanfx} and \EQ{muXg}.   }


%

\section{Bayesian Cramer-Rao Bound}\LABSEC{BCRB}

In this section we focus on a particular node of the grid, say $\vect{x}_{g_i}^{[t]}$. The GP provides us with a Bayesian estimate of the true RSS at this grid node, $f_{g_i}^{[t]}$,
\begin{eqnarray} 
\widehat{f}_{g_i}^{[t]}&\sim&\gauss{\widehat{f}_{g_i}^{[t]}}{{\mu}_{g_i}^{[t]}}{{\sigma}_{g_i}^{2\,[t]}},
\end{eqnarray}
i.e., our GP approach obtains the mean, ${\mu}_{g_i}^{[t]}$, by \EQ{mGP} and the variance ${\sigma}_{g_i}^{2\,[t]}$ by \EQ{vGP}.  
We want to find the lower bound of the \ac{MSE} of $\widehat{f}_{g_i}^{[t]}$, $\E_{\vect{z},f_{g_i}}[(f_{g_i}^{[t]}-\widehat{f}_{g_i}^{[t]})^2]$. We obtain this bound by computing the \ac{BCRB} \cite{Van07} 
according to \cite{Waagberg17}
\begin{align}\LABEQ{CRB1}
\E_{\vect{z},f_{g_i}}[(f_{g_i}^{[t]}&-\widehat{f}_{g_i}^{[t]})^2] \\
&\geq\left(\E_{\vect{z},f_{g_i}}\left[\left(\frac{\partial}{\partial f_{g_i}}\ln p(\vect{z}^{[t]},f_{g_i}^{[t]})\right)^2\right]\right)\inv ={\sigma}_{g_i}^{2\,[t]}. \nonumber
\end{align}

We note that apart from the estimated variable, $f_{g_i}^{[t]}$, we have {three} more parameters, $\alpha^{[t]}$, $P^{[t]}$ and {$\vect{x}_0$}, that introduce errors in the estimation.  
In \cite{Waagberg17}, the authors develop a \ac{HCRB} for \ac{GPR} with deterministic hyper-parameters.
The authors conclude that a term must be added to the variance of the \ac{GPR} in \EQ{CRB1}, as shown below. This term is a function of the derivative with respect to the hyper-parameters of the mean of the \ac{GPR}. The result is obtained under the assumption that the estimates of the hyper-parameters are unbiased, or at least asymptotically unbiased. 

We can cast our solution as a  \ac{GPR} where the deterministic hyper-parameters of the mean are $\boldsymbol{\mu}_{P,\alpha,\vect{x}_0}^{[t]}=[\widehat{\mu}_P^{[t]}, \; \widehat{\mu}_\alpha^{[t]},{\widehat{\vect{x}}_0\trs}]\trs$. Hence we may apply, in a straightforward way, the result in \cite{Waagberg17} to develop a \ac{HCRB} where the hyper-parameters $\widehat{\mu}_\alpha^{[t]}$, $\widehat{\mu}_P^{[t]}$ and ${\widehat{\vect{x}}_0}$ are considered as deterministic. The expression for the \ac{HCRB} is
%
%
\begin{eqnarray}\LABEQ{HCRB}
\E_{\vect{z},f_{g_i}}[(f_{g_i}^{[t]}-\widehat{f}_{g_i}^{[t]})^2]&\geq&{\sigma}_{g_i}^{2\,[t]}+ \vect{g}_{g_i}^{[t]\;\top}{\matr{M}_{g_i}^{[t]}}\inv\vect{g}_{g_i}^{[t]},
\end{eqnarray}
where
\begin{align}
\vect{g}_{g_i}^{[t]}\;\;&=\;\;\frac{\partial}{\partial \boldsymbol{\mu}_{P,\alpha,\vect{x}_0}^{[t]}}\left(m_{{\vect{x}_{g_i}}}-\vect{m}_{\widehat{\matr{X}}}^{[t]\,\top}{\left(\matr{C}_{\widehat{\matr{X}}}^{[t]}\right)}\inv \vect{k}_{\vect{x}_{g_i}{\widehat{\matr{X}}}}^{[t]\,\top}\right), \\
\matr{M}_{g_i}^{[t]}\;\;&=\;\;\frac{\partial \vect{m}_{\widehat{\matr{X}}}^{[t]\,\top}}{\partial \boldsymbol{\mu}_{P,\alpha,\vect{x}_0}^{[t]}} 
{\left(\matr{C}_{\widehat{\matr{X}}}^{[t]}\right)}\inv
\frac{\partial\vect{m}_{\widehat{\matr{X}}}^{[t]}}{\partial \boldsymbol{\mu}_{P,\alpha,\vect{x}_0}^{[t]}}.
\end{align}
We point out that the derivatives are with respect to the hyper-parameters $\widehat{\mu}_\alpha^{[t]}$ and $\widehat{\mu}_P^{[t]}$ and not with respect to $\alpha^{[t]}$, $P^{[t]}$. 
We can show that  
{\begin{align}
\vect{g}_{g_i}^{[t]}&=
\begin{bmatrix}
1 \\
-10\log_{10}({d}_{g_{i}}) \\
{\frac{c}{d_{g_i}^2 } (\widehat{\vect{x}}_{0}-\vect{x}_{g_i})}
\end{bmatrix}
-
\nonumber \\
&-  \begin{bmatrix}
{\sf \vect{1}} &  -\widehat{\vect{q}}^{[t]} & {\matr{A}}\\
\end{bmatrix}\trs {\left(\matr{C}_{\widehat{\matr{X}}}^{[t]}\right)}\inv \vect{k}_{\vect{x}_{g_i},{\widehat{\matr{X}}}}^{[t]\,\top} +
\nonumber \\
&+  
 \begin{bmatrix}
0 \\
0 \\
\vect{m}_{\widehat{\matr{X}}}^{[t]\,\top} {\left(\matr{C}_{\widehat{\matr{X}}}^{[t]}\right)}\inv \matr{A}_1{\left(\matr{C}_{\widehat{\matr{X}}}^{[t]}\right)}\inv  \vect{k}_{\vect{x}_{g_i},{\widehat{\matr{X}}}}^{[t]\,\top}\\
\vect{m}_{\widehat{\matr{X}}}^{[t]\,\top} {\left(\matr{C}_{\widehat{\matr{X}}}^{[t]}\right)}\inv \matr{A}_2{\left(\matr{C}_{\widehat{\matr{X}}}^{[t]}\right)}\inv  \vect{k}_{\vect{x}_{g_i},{\widehat{\matr{X}}}}^{[t]\,\top}
 \end{bmatrix},
 \\
\matr{M}_{g_i}^{[t]}&= \begin{bmatrix}
{\sf \vect{1}} &  -\widehat{\vect{q}}^{[t]} & {\matr{A}}\\
\end{bmatrix}\trs {\matr{C}_{\widehat{\matr{X}}}^{[t]}}\inv \begin{bmatrix}
{\sf \vect{1}} &  -\widehat{\vect{q}}^{[t]} & {\matr{A}}\\
\end{bmatrix},
\end{align}
where
\begin{align}
c&=-10\mu_{\alpha}\log_{10}(e), \\
\matr{A}&=c \widehat{\matr{D}} (\matr{J}_{N\times2}\mbox{diag}(\widehat{\vect{x}}_{0})-\widehat{\matr{X}}) \in \mathbb{R}^{N\times2}, \\
\matr{A}_1&=-2\sigma_u^2 \mbox{diag}\{\widehat{\matr{D}}^2 (\vect{\rm \vect{1}} \widehat{\vect{x}}_{0}(1) - \widehat{\vect{X}}(:,1))\} \in \mathbb{R}^{N\times N}, \\
\matr{A}_2&=-2\sigma_u^2 \mbox{diag}\{\widehat{\matr{D}}^2 (\vect{\rm \vect{1}} \widehat{\vect{x}}_{0}(2) - \widehat{\vect{X}}(:,2))\} \in \mathbb{R}^{N\times N},
\end{align}}
and $\widehat{\vect{x}}_{0}(j)$ is the $j$th element of the vector $\widehat{\vect{x}}_{0}$ and $\widehat{\vect{X}}(:,j)$ represents the $j$th column of the matrix $\widehat{\vect{X}}$.


At this point, we emphasize that our GPR has hyper-parameters, $\alpha^{[t]}$ and $P^{[t]}$, with priors that depend on their own hyper-parameters, $\boldsymbol{\theta}^{[t]}=[\mu_\alpha^{[t]},\sigma^{[t]}_\alpha,\mu_P^{[t]},\sigma^{[t]}_P,{\widehat{\vect{x}}_0}]$. We are using the results in \cite{Waagberg17} for the latter hyper-parameters. Therefore, we are somehow assuming that the variance of the GPR already includes the uncertainties  of $\alpha^{[t]}$ and $P^{[t]}$ by averaging in a Bayesian way. In turn, the HCRB provides the overall effect of $\mu_\alpha^{[t]}$ and $\mu_P^{[t]}$ on the \ac{MSE} of the RSS.

\section{Gaussian processes for time-varying fields}\LABSEC{recursiveGP}

We also consider the possibility that the RSS field may vary with time.  For example, the \ac{EIRP}, the orientation of the antennas, the objects around the sensors, among others, may change with time, and they may affect the RSS at the locations of the sensors. Also, some sensors may become unavailable, or they can move and change their positions. 
One possibility to address this is to simply use the approach from the previous section only on the current data and ignore the rest. An alternative is to include all the past information but enforcing reduced influence of older data on the estimates. 


As explained in \SEC{mod}, the main interest here is not finding how the field evolves at every position with time, which is already dealt by other approaches \cite{Perez13}. Instead, we just want to estimate the \ac{RSS} at the grid positions, i.e., at a set of predefined locations. We propose to update the estimates at these locations with information from the new observations. 
We compute the posterior distribution at the grid nodes, which is then used as a prior of the \ac{RSS} for the next instant time. 
In order to emphasize the information in the more recent samples and ``forget'' the information from old ones, we inject a forgetting factor, $0<\lambda\leq1$. As a result, our solution is a linear combination of past and current information weighted by $\lambda$ and $1-\lambda$, i.e., 
\begin{align}\LABEQ{postRGP}
p\left(\vect{f}_{g}^{[t]}|\matr{X}_{g},{\widehat{\matr{X}}}^{[1:t]},\vect{z}^{[1:t]}\right)&= \gauss{\vect{f}_{g}^{[t]}}{\boldsymbol{\mu}_g^{[t]}}{\boldsymbol{\Sigma}_g^{[t]}},
\end{align}
where 
\begin{align}
\boldsymbol{\mu}_g^{[t]}&= \vect{m}_{\matr{X}_g}^{[t]}+ (1-\lambda) \boldsymbol{\mu}_{prior}^{[t]} + \lambda \boldsymbol{\mu}_{post}^{[t]} , \LABEQ{meanRGP}
\\
\matr{\Sigma}_g^{[t]}&=\matr{K}_{\matr{X}_g}^{[t]}- \Big((1-\lambda) \matr{\Sigma}_{prior}^{[t]} + \lambda\matr{\Sigma}_{post}^{[t]}\Big),
\LABEQ{covRGP}
\end{align}
and
\begin{align}
\boldsymbol{\mu}_{post}^{[t]}&=\matr{K}_{\matr{X}_g,{\widehat{\matr{X}}}}^{[t]}{\left(\matr{K}_{\widehat{\matr{X}}}^{[t]}+\boldsymbol{\Sigma}_\noise^{[t]}\right)}\inv \left(\vect{z}^{[t]}-\vect{m}_{\widehat{\matr{X}}}^{[t]}\right), 
\\
\matr{\Sigma}_{post}^{[t]} &= \matr{K}_{\matr{X}_g,{\widehat{\matr{X}}}}^{[t]}{\left(\matr{K}_{\widehat{\matr{X}}}^{[t]}+\boldsymbol{\Sigma}_\noise^{[t]}\right)}\inv  \matr{K}_{\matr{X}_g,{\widehat{\matr{X}}}}^{[t]\,\top}, 
\\
\boldsymbol{\mu}_{prior}^{[t]}&=
\boldsymbol{\mu}_g^{[t-1]}-\vect{m}_{\matr{X}_g}^{[t-1]},
\\
\boldsymbol{\Sigma}_{prior}^{[t]}&= \matr{K}_{\matr{X}_g}^{[t-1]}-
\matr{\Sigma}_g^{[t-1]}.
\end{align}
Due to the dynamic nature of the scenario, a new estimation of $\alpha$ and $P$ is computed at every time instant following \EQ{meanPalpha}. Note that we are ensuring that the covariance matrix in \EQ{covRGP} remains positive definite due to the linear combination weighted by $\lambda$ and $1-\lambda$. 
Our approach is summarized in \ALG{recGP}, which we refer to as recursive Gaussian process (rGP)-based algorithm.

\begin{algorithm}[htb]
\begin{algorithmic}
\STATE{\bf Initialization.} Compute the posterior probability at the grid nodes for $t=0$ by executing \ALG{staticGP} 
and set
$\boldsymbol{\mu}_{prior}^{[0]}=\boldsymbol{\mu}_g^{[0]}$,  
$\matr{\Sigma}_{prior}^{[0]} =\boldsymbol{\Sigma}_g^{[0]}$.
\FOR {$t=1, 2, ...$}
\STATE
Compute the posterior distribution at the grid nodes at time instant $t$, i.e., $p\left(\vect{f}_{g}^{[t]}|\matr{X}_{g},{\widehat{\matr{X}}}^{[1:t]},\vect{z}^{[1:t]}\right)$ as in \EQ{postRGP}. 
\ENDFOR 
\end{algorithmic}
\caption{Recursive GP (rGP) formulation}\LABALG{recGP}
\end{algorithm}

Note that when $\lambda= 1$, we forget all the prior knowledge about the previous data and compute the mean and covariance matrix of the GP with just the current data, yielding the approach developed in \SEC{GP} \cite{Santos17b}. 


\section{Experimental results}\LABSEC{sim}

In this section we include results that illustrate the performance of the proposed approach. First, we explain the setup.

\subsection{Simulated scenario}\LABSSEC{simscenario}

The setting is similar to the one used in \cite{Santos17b}. We simulated an area of  500 m $\times$ 500 m where a single transmitter is placed at the center of the area. We considered a fixed and uniform grid with $M=1088$ nodes, where we want to estimate the RSS. 
We randomly placed $N=218$ sensors within the area, and their RSS measurements were generated according to \eqref{measurements} with $\hat{d}_i^{[t]}$ replaced by ${d}_i^{[t]}$ and with $\vect{u}^{[t]}$ removed, $\alpha^{[t]}=3.5$ and $P^{[t]}=-10$ dBm. The rest of the parameters were set to  $\sigma_w=\sqrt{7}$ dB, $\sigma_v=\sqrt{10}$ dB, $\rho_u=200$ mdB, and $D_{corr}=50$ m. {We assumed that the values of the parameters $\sigma_v$ and $D_{corr}$ are not available and that they would be estimated from the GP. }

One example of this scenario is shown  in \FIG{sysmod}, where we represent the grid nodes with red squares, the transmitter with a green triangle and the sensors with circles. The different values of RSS at the sensors are represented with a scale of colors where yellow means the highest  and blue the lowest values. 

\begin{figure}[htb]
\begin{center}
\includegraphics[width=3.5in]{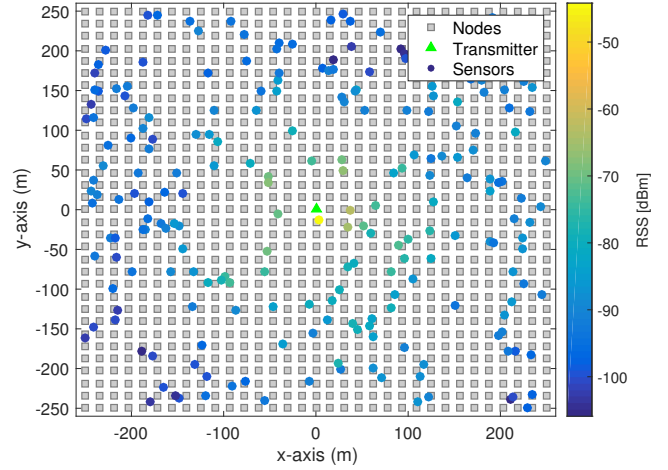}
\caption{A graphical representation of the space of interest with the locations of the transmitter, nodes, and sensors.}
\LABFIG{sysmod}
\end{center}
\end{figure}

To quantify the error of the performance, we used the mean squared error (MSE). The metric was computed for each time instant as
\begin{align}
\mbox{MSE}^{[t]}&=&\frac{1}{M}\sum_i \left(\mu_{g_i}^{[t]}-f_{g_i}^{[t]}\right)^2, \;\;\forall i=1\cdots M, 
\end{align}
where $f_{g_i}$ is the true value of the RSS at the $i$th node and $\mu_{g_i}^{[t]}$ is its estimate. 

\subsection{Error in the location}\LABSSEC{simLoc}

In this subsection we show the performance of our GP algorithm for static fields considering the three different cases explained in \SEC{mod}, i.e.,
\begin{enumerate}
\item Case 1: Using the true sensor locations and the model in \EQ{measurementsNoNoise}.
\item  Case 2: Using the sensor locations with errors and accounting for the errors as in \eqref{measurements}.
\item Case 3: Using the sensor locations with errors as if they were accurate and following \eqref{mod3}.
\end{enumerate}

In \FIG{MSECasesv3} we show the MSEs for these three cases. Clearly, when there is no error in the location (Case 1), we obtain the lowest error.  On the other hand, when considering locations with errors (Cases 2 and 3), the performance is just slightly deteriorated in comparison to Case 1. Specifically, our proposal of introducing one additional source of error to model the error of the user location (Case 2) improves the traditional approach of ignoring this source of error (Case 3).



\begin{figure}[htb]
\centering
\includegraphics[width=3.5in]{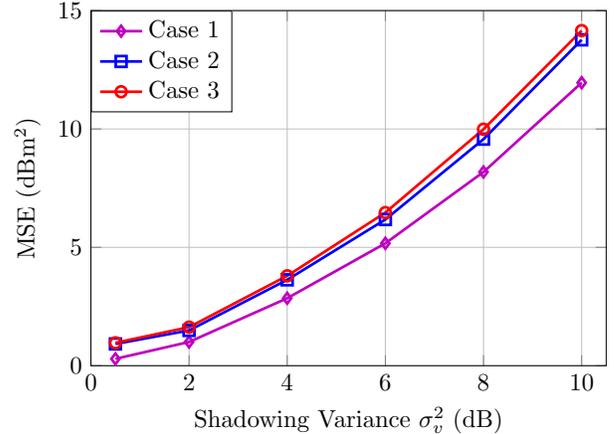}
\caption{\small Mean square error for varying shadowing variance and $\rho_u=200$ mdB ($\sigma_\epsilon=13.1576$). } \LABFIG{MSECasesv3}
\end{figure}

\subsection{Static GP}
In this subsection we analyze the performance of the sGP approach for static fields. One example of such setting is depicted in \FIG{sysmod}. 


We first estimated \newt{the transmitter location according to \EQ{loc_est}} and the path loss exponent and the transmitter power following \EQ{meanPalpha} and \EQ{varPalpha}. \newt{Then, these parameters were refined as explained in \SSEC{refine}.} To show the robustness of our method for estimating the mean of the path loss exponent and transmitted power, we introduced a high error in the location of the closest sensor to the transmitter. The obtained values of the means are shown in \TAB{estimationPalpha}, and they are close to the true values. \newt{Further, we show estimation results for different positions of the transmitter.} 
%


\begin{table}[htb]
\caption{Estimates of the path loss exponent, transmitted power and location of the transmitter. }
\includegraphics[width=3.5in]{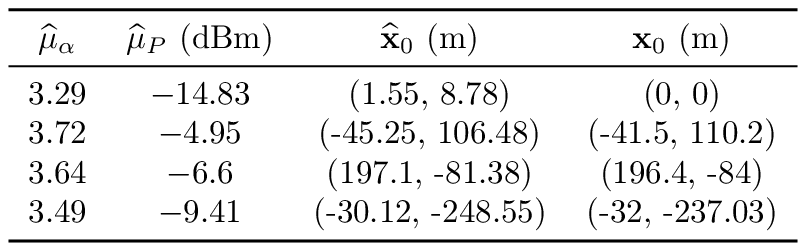}
 \LABTAB{estimationPalpha}
\end{table} 



We also include \FIG{alphaPest} where we compare the estimated and true \ac{RSS} along different distances from the transmitter. The figure also displays the noisy measurements of the \ac{RSS} of the sensors (plotted with circles). 
The results show that the estimated values (red dashed line) are approximately the same as the true ones (black solid line), and in agreement with the results from \TAB{estimationPalpha}.  {Note that the closest measurement to the transmitter (whose RSS is around $-$50 dBm) did not affect negatively the estimates of $\alpha$ and $P$ because of the precaution we took with \EQ{meanPalphaChi}. }

\begin{figure}[htb]
\begin{center}
\includegraphics[width=3.5in]{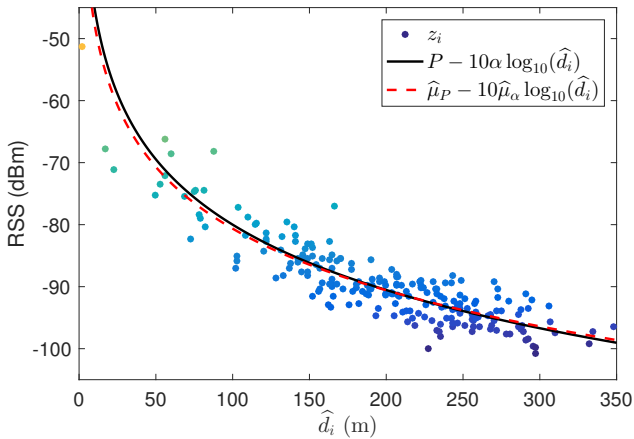}
\caption{Measured powers by the sensors (circles), estimated RSS (dashed red), and true RSS (solid black) as functions of distance from the transmitter.  }\LABFIG{alphaPest}
\end{center}
\end{figure}

The results of the RSS field estimates are shown in \FIG{GP}. We represent the RSS measurements of the sensors with red circles and the estimated mean of the posterior distribution of the GP over the coverage area with blue solid surface. The graph demonstrates that the GP smoothly approximates the RSS field.  


\begin{figure}[htb]
\begin{center}
\includegraphics[width=3.5in]{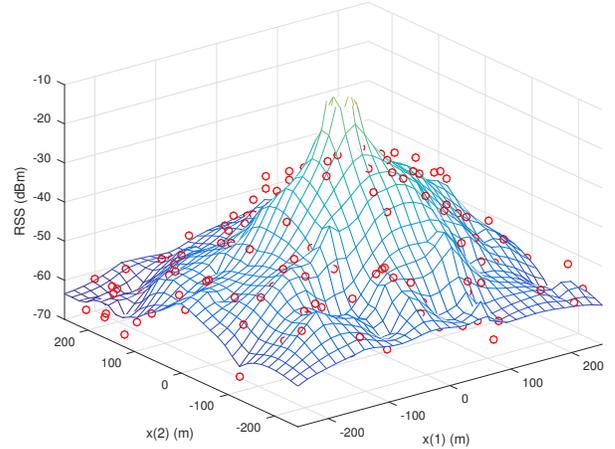}
\end{center}
\caption{\small Mean of the estimation (solid) and observations ($\circ$). }
\LABFIG{GP}
\end{figure}

Finally, in \TAB{estimatedhyp} we present the true and estimated values of the hyperparameters of the kernel. 

\begin{table}[htb]
\caption{True and estimated hyperparameters of the kernel. }
\includegraphics[width=3.5in]{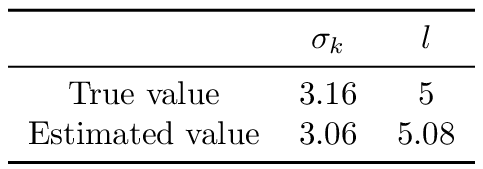}
 \LABTAB{estimatedhyp}
\end{table} 


\subsection{Recursive GP}

Here we provide some results with the recursive GP. 
In \FIG{MSErecursive}, we display results from two different scenarios:
1) an intermittent setting where 20\% of the sensors are unavailable at each time instant and 2) a setting where the sensors are moving from their previous positions. The figure shows plots of MSEs for different shadowing variances and at two time instants ($t=1$ and $t=10$). The results were averaged over 100 different experiments and compared 
to the \ac{OKD} technique \cite{Molinari15,Montero15}, applied at $t=1$.
As expected, the recursive approach gets better estimates with time. Also as expected, the error of the rGP method with intermittent data is slightly higher than with data of moving sensors. Note that when we use intermittent data at $t=1$, the performance is not as good as when we use data from moving sensors because there are less available measurements.  


\begin{figure}[htb]
\centering
\includegraphics[width=3.5in]{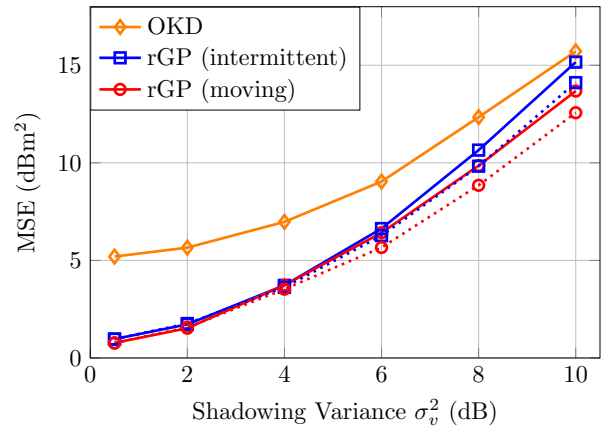}
\caption{\small Mean square error for $\rho_u=200$ mdB and for $t=1$ (solid) and $t=10$ (dotted). } \LABFIG{MSErecursive}
\end{figure}

Finally, in \FIG{CRB}, we illustrate the time variability of the RSS at a specific location when {\it dynamic transmitter powers} are considered. We changed the value of $P^{[t]}$ two times during the observation interval so that the RSS varied as shown by the green dashed line. The blue solid line represents the mean of the estimated posterior obtained from the GP following \EQ{meanRGP}, and the gray shadowed error bars are bands  around the mean whose widths are equal to three times the square root of HCRB deviation in \EQ{HCRB}. The applied value of $\lambda$ was 0.5.

\begin{figure}[htb]
\centering
\includegraphics[width=3.5in]{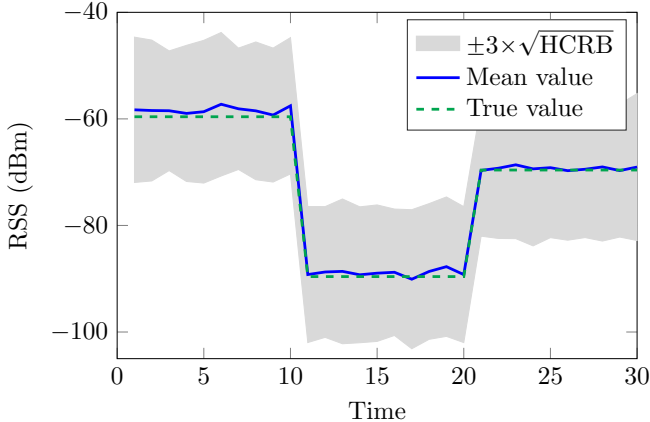}
\caption{\small Mean of the estimation (blue solid) and $\pm3\times \sqrt{HCRB}$ (filled), given by \EQ{HCRB}, and true RSS (green dashed)  along time. }
\LABFIG{CRB}
\end{figure}

\section{\newt{Experimental results with real data}}
\LABSEC{New}
\newt{For testing our method with real data, we used a GSM dataset from a study reported on \cite{Chakraborty15}. The measurements were collected on the campus of Stony Brook University with moving Nexus 5 smartphones, which were running Android.  The dataset was collected during random days over a month, and the total number of measurements was 6437. The location of the base station was perfectly known. 
}


\begin{figure}[htb]
\centering
\includegraphics[width=3.5in]{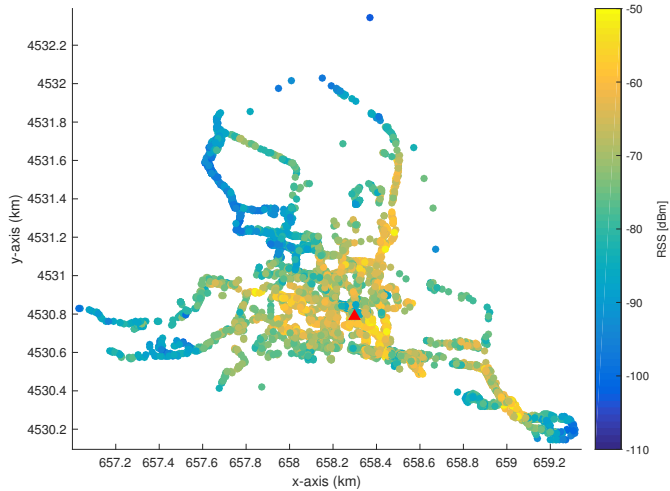}
 \caption{{A graphical representation of the dataset from \cite{Chakraborty15}. The location of the transmitter is marked with a red triangle (\textcolor{red}{\scriptsize{$\blacktriangle$}}) and the locations of the sensors with circles. The colors of the circles reflect the RSS.}}
 \LABFIG{RSSayonOmni}
\end{figure}

\newt{In testing our method with these data, we randomly divided them into two groups, training (50\%) and testing (50\%) data.  
Note that, unlike the square uniform grid we used for the synthetic dataset in \SEC{sim},  we had a nonuniform one because we placed the grid points at the locations where the test data were acquired. 
We ran our algorithm with $\rho_u=1140$ mdB and $\sigma_w=\sqrt{7}$ dB. Note that we increased the value of $\rho_u$ in comparison to the one in \SEC{sim} because the distances were of the order of 100 m. Then it was logical to increase the standard deviation of the actual error in distances to $\sigma_d=75$ m, which was equivalent to setting $\rho_u=1140$ mdB.}

\newt{First, we estimated the path loss exponent and the transmitter power following \EQ{meanPalphaChi}. 
The results are shown in the first two columns of \TAB{estimationPalphaReal}. The estimated values of the hyperparameters of the kernel are given in the next two columns. The MSE of  the RSS at the nodes is given in the last column. }

%
%
%

\begin{table}[htb]
\caption{Estimated values of the path loss exponent, transmitted power, hyperparameters of the kernel and the obtained MSE when estimating the RSS at unavailable nodes in the test set. }
\includegraphics[width=3.5in]{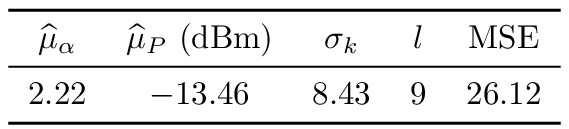}
 \LABTAB{estimationPalphaReal}
\end{table} 

Note that in a real-world scenario, the model in (\ref{measurements}) does not perfectly fit the measurements.  Specifically, the RSS can also be affected by obstacles that obstruct the line of sight, the path loss exponent might depend on the position where a measurement is taken, or the antenna gain diagram might not be perfectly omnidirectional, among others. All these factors cause the value of the MSE to increase in comparison to the MSEs obtained from simulated measurements where basically we have a perfect model match.

%


\section{Conclusions}\LABSEC{conclusion}

In this paper, we use crowdsourcing to solve the problem of \ac{RSS} estimation of a possibly time-varying field. We rely on measurements provided by different and inexpensive sensors that are randomly placed within an area of interest. We developed a Bayesian framework based on \ac{GP} and a model with several unknown parameters. The unknown path loss exponent and transmitted power were modeled as Gaussian variables. The hyper-parameters of their Gaussian distributions were estimated from the data. We also assumed that the user locations were not perfectly known, which introduced an additional source of error in the model. Further, the location of the transmitter was unknown too, and it was estimated from the data, which had its own error. We also addressed the problem where the field may vary with time. In our solution we used a forgetting factor which determines the relevance of previous and current data. In all our solutions, the needed memory of the algorithm is fixed and a function of the number of nodes, and it is independent of the number of sensors. Finally, we derived the \ac{HCRB} of the estimated parameters, and we showed the performance of our approach with experimental results on synthetic \newt{ and real} datasets.



\appendix[Model of error due to location estimates]\LABAPEN{apen1}
We start by taking the logarithm of \eqref{location}
\begin{align}
\log_{10}\left(\widehat{{d}}_i^{[t]}\right) &= \log_{10}\left(d_i^{[t]}+\epsilon_d\right) =  \log_{10}\left(d_i^{[t]}\left(1+\frac{\epsilon_d}{d_i^{[t]}}\right)\right)\nonumber\\ 
&=\log_{10}\left(d_i^{[t]}\right)+\log_{10}\left(1+\frac{\epsilon_d}{d_i^{[t]}}\right).
\end{align}
Let $x={\epsilon_d}/{d_i^{[t]}}$, which is a variable that takes small values. Then we expand $f(x)=\log_{10}(1+x)$ by Taylor expansion around $x=0$ and get
\begin{align}
f(x)&=\log_{10}(1+x)|_{x=0}+\frac{\log_{10}(e)}{1+x}|_{x=0}x+R \nonumber\\
&=\log_{10}(e)x+R=\frac{\log_{10}(e)\epsilon_d}{d_i^{[t]}}+R,
\end{align}
where $R$ is the residual \newt{and $e=\exp(1)$}. Thus,
\begin{align}
\log_{10}\left(\widehat{{d}}_i^{[t]}\right) \approx \log_{10}\left(d_i^{[t]}\right)+\frac{\log_{10}(e)\epsilon_d}{d_i^{[t]}},
\end{align}
or
\begin{align}
\label{eq:app}
\log_{10}\left({{d}}_i^{[t]}\right) \approx \log_{10}\left(\widehat{d}_i^{[t]}\right)-\frac{\log_{10}(e)\epsilon_d}{d_i^{[t]}}.
\end{align}
Replacing \eqref{eq:app} in \EQ{measurementsNoNoise}, we obtain that $u_i^{[t]}$ in \eqref{measurements} is 
\begin{align}\LABEQ{u}
u_i^{[t]} = -\frac{10\alpha^{[t]}\log_{10}(e)}{\widehat{d}_i^{[t]}-\epsilon_d}\epsilon_d.
\end{align}
If we assume that $\widehat{d}_i^{[t]}>>\epsilon_d$ and $\epsilon_d\sim \gauss{\epsilon_d}{0}{\sigma_d^2}$, we can write 
\begin{align}
\label{eq:u_t}
u_i^{[t]}\sim\gauss{u_i^{[t]}}{0}{\frac{\left(10\alpha^{[t]} \log_{10}(e)\right)^2}{\widehat{d}_i^{2[t]}}\sigma_d^2},
\end{align}
Thus, for the standard deviation of $u_i^{[t]}$, we have $\sigma_u^{[t]}=\rho_u^{[t]}/\widehat{d}_i^{[t]}$, where
\begin{align}
\rho_u^{[t]}={10\alpha^{[t]}\sigma_d\log_{10}(e)}, 
\end{align}
which is what we have in \EQ{erroru}.


%

%


%
%

\ifCLASSOPTIONcaptionsoff
  \newpage
\fi



%
%
%

\bibliographystyle{IEEEtran}
\bibliography{allBib,bounds,dynamics4,rssMeasurement.bib}

\end{document}